\documentclass[a4paper,10pt]{article}
\usepackage[utf8]{inputenc}

\usepackage{graphicx}
\usepackage{textcomp}
\usepackage{subfigure}
\usepackage{amsmath,amssymb}
\usepackage{authblk}
\usepackage{ctable}

\usepackage[hmargin=2.5cm,vmargin=2.3cm]{geometry}
\title{Chaotic behavior of the Compound Nucleus, open Quantum Dots and other nanostructures}
\author[1,2]{M. S. Hussein}

 \author[3] {J. G. G. S Ramos}

\affil[1]{Instituto de Estudos Avan\c cados, Universidade de S\~ao Paulo, C.P. 72012, 05508-970, S\~ao Paulo, SP, Brazil.}

\affil[2]{Departamento de Fisica Matem\'atica, Instituto de F\'isica da Universidade de S\~ao Paulo, C. P.  66318, 05314-970, S\~ao Paulo, SP, Brazil.}

\affil[3]{Departamento de F\'isica, Universidade Federal da Para\'iba, 58051-970, Jo\~ao Pessoa, PB, Brazil.}
 
\begin{document}

 \maketitle
\begin{abstract}
It is well established that physical systems exhibit both ordered and chaotic behavior. The chaotic behavior of nanostructure such as open quantum dots has been confirmed experimentally and discussed exhaustively theoretically. This is manifested through random fluctuations in the electronic conductance. What useful information can be extracted from this noise in the conductance?  In this contribution we shall address this question. In particular, we will show that the average maxima density in the conductance is directly related to the correlation function whose characteristic width is a measure of  energy- or applied magnetic field- correlation length. The idea behind the above has been originally discovered in the context of the atomic nucleus, a mesoscopic system. Our findings are directly applicable to graphene.
\end{abstract}
%
%
\section{Introduction}
\label{intro}
It is by now well proven that complex quantum systems exhibit fluctuation behaviors which can be described using statistical methods. The compound nucleus is one such system.
One resorts to energy averaging or ensemble averaging to obtain the averaged cross section. Further, the case of overlapping resonances, requires the study of the correlation function in order to obtain information about the lifetime of the compound nucleus. The resulting theory of compound nucleus hinges on the Hauser-Feshbach expression for the average cross section \cite{HF}, and the Ericson  correlation function \cite{Ericson}. There is a vast body of literature on the compound nucleus theory. The techniques employed found their way to other physical systems. In particular nanostructures, such as quantum dots, have been discussed using the compound nucleus theory techniques.

In nano systems, one has a control on their behavior through the changes of external parameters. Further, in the case of an open QD, a nano system of zero diminution, where the electrons enter and exit the dots through the leads, one has an analog computer of the compound nucleus, since resonances in the electron conductance are in general overlapping and resemble very much the behavior of complex chaotic systems. Other nano systems of higher dimensions are fabricated and studied. Quantum wires have a dimension of 1 while graphine,  has a dimension of 2. In the following, we study the statistical properties of QD, depicted in Fig. [1]. It consists of a chaotic quantum dot connected to two electron reservoirs via ideal leads with a number of open channels. Electron reservoirs have temperature $ T $ and electrochemical potentials $ \mu_{1} $ and $ \mu_{2} $, generating a potential difference $ e V $. The injected electrons have a finite probability of tunneling for entrance and
  exit due to the presence of barriers $ \Gamma_{1} $ and $ \Gamma_{2} $. The QD can couple the spin and orbital degrees of freedom through relativistic corrections owing to the band structure and/or asymmetries in interfaces of heterostrutures.

\begin{figure}
\centering
\includegraphics[width=12cm,clip]{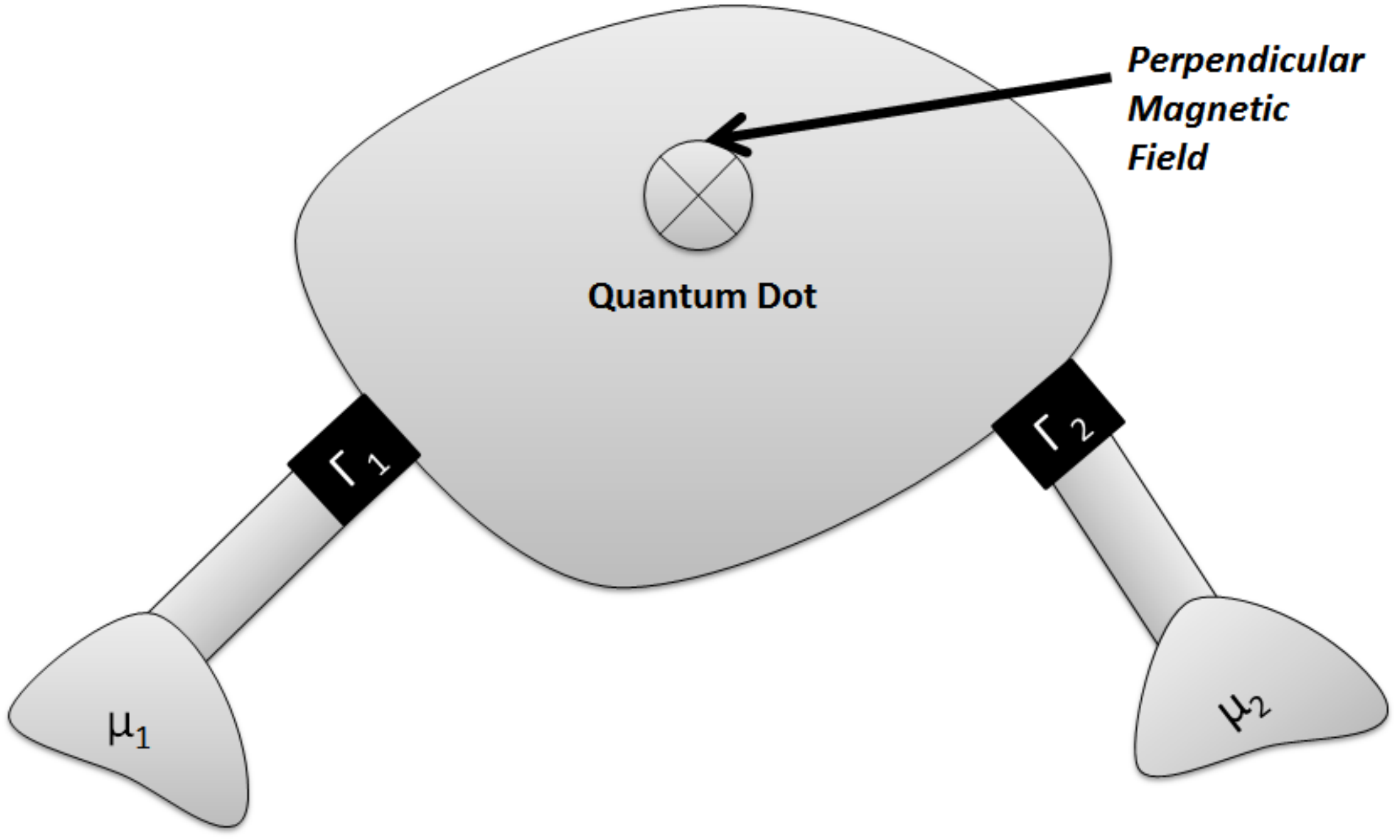}
\caption{A schematic diagram showing a quantum dot coupled to two leads.}
\label{Fig.1}       
\end{figure}

There is an extensive literature devoted to the study of the statistical properties of the electronic conductance in ballistic open quantum dots (QDs) containing a large number of electrons.\cite{beenakker97} In such systems, it is customary to assume that the underlying electronic dynamics is chaotic to statistically describe the electron transport properties using the random matrix theory (RMT) \cite{weiden1, weiden2, weiden3} and the Landauer conductance formula\cite{Land, Butti} Within this framework, the conductance fluctuations are universal functions that depend on the quantum dot symmetries, such as time-reversal, and on the number of open modes N connecting the QD to its source and drain reservoirs.
In the semiclassical limit of large N, the transmission statistical fluctuations are accurately modeled by Gaussian processes. In practice, it has been experimentally observed \cite{5} and theoretically explained \cite{6} that, even for small values of N and at very low temperatures, dephasing quickly brings the QD conductance fluctuations close to the Gaussian limit.
The electronic conductance in open ballistic QDs exhibits random fluctuations as an external parameter, such as a magnetic field B or an applied gate voltage Vg, is varied. So far, the majority of the studies concentrate on the study of conductance distributions, which usually demand very large experimental statistical sampling to allow comparison with theory. This invites one the ask whether something useful can be extracted from a direct analysis of a single conductance curve. Inspired by the formal analogy between conductance and compound-nucleus Ericson fluctuations \cite{Alhassid} we show that the answer is positive. More specifically, inspired by results obtained about 50 years ago in nuclear physics,\cite{Brink}, (see also \cite{Bizzeti, Bonetti}) we calculate the average density of maxima in the conductance and show its relation with the conductance correlation function. Using the random matrix theory prediction for the latter, we find a new universal observable in the ballistic conductance of open quantum dots. The content of this contribution are based on two recent publications, \cite{RBHL11, BHR13}.

\section{Ericson Fluctuations}

The average value of fluctuation quantities of a physical system are calculable with available theories. This is the case of the average compound nucleus cross section, which is given by the Hauser-Feshbach theory. The same holds for open Quantum Dots. A more stringest measure of the nature of these fluctuating observables can be best quantified with the aid of the correlation function. The resulting function is Lorentzian in shape as demonstrated by Ericson \cite{Ericson}. Consider the conductance in QD. It is given by the Landauer-B\"uttiker formula \cite{Land, Butti},

\begin{equation}
G = \frac{2e}{h} T = \frac{2e}{h} Tr(t^{\dagger}t)
\end{equation}
where $T$ is the dimensionless conductance given in terms of the transmission matrix, which appears in the S-matrix as \cite{mello},

\begin{equation}
\mathbf{S} = \left(
  \begin{array}{cc}
  r &t \\
  t^{\prime} & r^{\prime}\\
  \end{array} \right)
\end{equation}

The usual decomposition of $T$ into an average part plus a fluctuation part is made,
\begin{equation}
T = \left <T \right >+ T^{fl}
\end{equation}

Then the correlation function is given by
\begin{equation}
C(\delta Z) = \langle{T^{fl}(Z + a \delta Z) T^{fl}(Z - b \delta Z)\rangle},
\end{equation}
where $Z$ can be the energy or any other quantity that may modify the Hamiltonian. The above correlation function does not depend on the choice of $a$ and $b$, provided that $a + b = 1$. In nuclear physics the only quantity that is available for change and study is the energy. In QD, on the other hand, besides the energy, one may apply a magnetic field or change the shape both of which are treated as external parameters. For the case $Z = E$, the result of the average above is given by Ericson correlation function,

\begin{equation}
C(\varepsilon) = \frac{\langle{T^{fl}T^{fl}\rangle}}{1 + (\frac{\varepsilon}{\gamma})^2}
\end{equation}

In the case of a variation of an external parameter, $X$, Efetov \cite{Efetov, Caio}, has shown that the correlation function is a squared Lorentzian,

\begin{equation}
C(\delta X ) =  \frac{\langle{T^{fl}T^{fl}\rangle}}{[1 + (\frac{\delta X}{X_c})^2]^2}
\end{equation}

In the above equations, the average $\langle{T^{fl}T^{fl}\rangle}$ is the variance. In the case of the compound nucleus, this quantity is the square of the Hauser-Feshbach cross section. Further, if direct reactions are present, the generalized correlation function for the transition from channel $c$ to channel $c^{\prime}$ is then given by,

\begin{equation}
C_{cc^{\prime}}(\varepsilon) = \frac{2\sigma^{d}_{cc^{\prime}} \sigma^{fl}_{cc^{\prime}} + (\sigma^{fl}_{cc^{\prime}})^2}{1 + (\frac{\varepsilon}{\gamma})^2}
\end{equation}
In the above formulae, the correlation lengths are the inverse of the life time, $\gamma$, of the underlying resonant system, and the characteristic length $X_c$ of the fluctuations arising from varying the external parameter having a more complicated structure. It has been proven that $\gamma$ is related to the density of states of the system, $\rho = 1/D$,  , where $D$ is the average resonance spacing, through the formula,

\begin{equation}
\gamma = \frac{D}{2\pi} Tr{ P}
\end{equation}

where $P$ is the transmission probability (or the tunneling probability) matrix. This is usually calculated with optical potential in the nuclear case. In the case of QD, the $P$'s are coefficients that are varied externally. In the case of $P =1$, we have the interesting sum rum for QD,

\begin{equation}
\gamma = \frac{D}{2\pi} N
\end{equation}
where $N$ is the number of open channels. \\

Through the above sum rule, one can determine the density of states, once the correlation analysis is performed and $\Gamma$ is extracted. A simple formula for $X_c$ is more difficult to obtain. However it can be shown that $X_c$ depends on the number of open channels, as $X_c \sim \sqrt{N}$ \\

A general correlation function for QD can be derived in the case of a tunneling probability, $\Gamma$, smaller than one (all the above results assume $\Gamma = 1$). One finds \cite{BHR13}, using the stub model \cite{brouwer},

\begin{equation}
C(\varepsilon)= \langle{T^{fl}T^{fl}\rangle} \left[\frac{3\Gamma(2 - \Gamma) - 2}{1 + (\varepsilon/\Gamma)^2} + \frac{4[1 + \Gamma (\Gamma -1)]}{[ 1 + (\varepsilon/\gamma)^2]^2}\right]
\end{equation}

and,
\begin{equation}
C(\delta X) = \langle{T^{fl}T^{fl}\rangle}\left[\frac{2\Gamma(1 - \Gamma)}{1 + (\delta X / X_c)^2} + \frac{2 + \Gamma (3\Gamma - 4)}{[1 + (\delta X/ X_c)^2]^{2}} \right]
\end{equation}

The above correlation functions reduce to the ones in Eqs, (5) and (6), in the limit of fully open QD. Further, the energy correlation function, $C(\varepsilon)$ above exhibits anti-correlation (negative correlation function). In fact in the limit of very small tunneling ($\Gamma \rightarrow 0$), one has,

\begin{equation}
C(\varepsilon) = \langle{T^{fl}T^{fl}\rangle}\frac{1 - (\varepsilon/ \gamma)^2}{[1 + (\varepsilon/\gamma)^2]^2}
\end{equation}

Ericson-type analysis has been extensively used in compound nuclear reaction studies. Further, it has been employed by researchers in other areas to characterize chaotic scattering systems in general, and has become an integral part of the theoretical investigation of complex chaotic systems.

\section{Number of maxima method}


The un-averaged cross section in compound nucleus reactions or electron conductance in QD, exhibits energy fluctuations which show the characteristics of noise. One may wonder whether something useful can be extracted from such a chaotic function.
Since the underlying S-matrix is random in nature, one could trace some features of the fluctuating observable to this randomness. In the following we calculate the average density of maxima in the un-averaged
, fluctuating, observable and show that it is directly related to the correlation function. This added information sheds more light on the nature of chaotic scattering systems.\\

The condition that the random observable has a maximum at energy $E$ is,
\begin{equation}
\tau_{fl}(E) > \tau_{fl}(E \pm \varepsilon_0).
\end{equation}
We call, $\tau_1 = T_{fl}(E), \tau_2 = T_{fl}(E +\varepsilon_0), \tau_3 = T_{fl}(E - \varepsilon_0)$, with,
\begin{equation}
\tau_1 > \tau_2 > \tau_3. \label{123}
\end{equation}
We now introduce the joint probability distribution of the three random numbers, $\tau_1, \tau_2,$ and $\tau_3$. We do this by first recalling the Gaussian distribution of a single variable, $x$,
\begin{equation}
P(x) = \frac{1}{\sqrt{2\pi\sqrt{\langle{x^2}\rangle}}}\exp{[-\frac{1}{2}\frac{x^2}{\langle{x^2\rangle}}]}
\end{equation}
where $\langle{x^2\rangle}$ is the variance or second moment of the distribution,
\begin{equation}
\langle{x^2\rangle} = \int_{-\infty}^{\infty} x^2dxP(x) \label{entropy1}
\end{equation}
Of course $P(x)$ is normalized by construction,
\begin{equation}
\int_{\infty}^{\infty} dx P(x) = 1.\label{entropy2}
\end{equation}
The above Gaussian distribution can be obtained from the principle of maximum information entropy subjected only to constraints given by conditions Eqs. (\ref{entropy1}, \ref{entropy2}), namely,
\begin{equation}
\delta\left[S + \lambda_1<x^2> + \lambda_2<1>\right] = 0
\end{equation}
where the information entropy $S$, is given by,
\begin{equation}
S = -\int dx P(x)\ln P(x).
\end{equation}
Performing the variational calculation of Eq. (ref{entropy3}), we get,
\begin{equation}
[\ln P +1 +\lambda_{1} x^2 + \lambda_2]\delta P = 0,
\end{equation}
which yields,
\begin{equation}
P(x) = e^{1 - \lambda_2 }\exp{[-\lambda_{1}x^2]}
\end{equation}
The generalization of the above to the distribution of the components of a 3-vector is now straightforward,
\begin{equation}
P(\tau_1, \tau_2, \tau_3) = \frac{1}{(2\pi)^{3/2}D^{1/2}}\exp{\left[-\frac{1}{2}X^{T}AX\right]},
\end{equation}
where the vector $X^{T} =(\tau_1, \tau_2, \tau_3)$, is its transpose of $X$, and $D = det C$, with C being the correlation matrix, whose ${kj}$ element is just the S-matrix
two point correlation function,
\begin{equation}
 \mathbf{C} = \left(
  \begin{array}{ccc}
  \langle{\tau_1\tau_1}\rangle & \langle{\tau_1\tau_2}\rangle & \langle{\tau_1\tau_3}\rangle\\
  \langle{\tau_2\tau_1}\rangle & \langle{\tau_2\tau_2}\rangle & \langle{\tau_2\tau_3}\rangle\\
  \langle{\tau_3\tau_1}\rangle & \langle{\tau_3\tau_2}\rangle & \langle{\tau_3\tau_3}\rangle\\
  \end{array} \right)
\end{equation}
\begin{equation}
C_{kj} = \langle{\tau_k\tau_j}\rangle = |\frac{T_{fl}}{1 + i(k -j) \frac{\varepsilon_0}{\gamma}}|^2.
\end{equation}
The matrix $A = C^{-1}$.
With the above, we can calculate the average number of maxima, through integration over the three variables, under the constraint of Eq. (\ref{123}),
\begin{equation}
\langle{\rho_E}\rangle = \frac{2}{\varepsilon_0}\int_{-\infty}^{\infty}d\tau_2\int_{-\infty}^{\tau_2}d\tau_1\int_{-\infty}^{\tau_1}d\tau_3F(\tau_1, \tau_2, \tau_3)
\end{equation}
Integrating the above equation yields,
\begin{equation}
\langle{\rho_E}\rangle = \frac{1}{\pi \varepsilon_0} \tan^{-1}\sqrt{4\frac{C(0) - C(\varepsilon_0)}{C(0) - C(2\varepsilon_0)} -1}
\end{equation}
Using the excplicit form of the correlation function, $C(\varepsilon) = (T_{fl})^2/(1 + \varepsilon^{2}/\gamma^2)$, we get,
\begin{equation}
\langle{\rho_E}\rangle = \frac{1}{\pi \varepsilon_{0}}\tan^{-1}\sqrt{3\frac{\varepsilon_{0}^2}{\gamma^2 + \varepsilon_{0}^2}}
\end{equation}

In the limit of $\varepsilon_0 = 0$, the above relation reduces to,
\begin{equation}
\langle{\rho_E}\rangle =\frac{\sqrt{3}}{\pi\gamma} = \frac{0.55}{\gamma}
\end{equation}
Therefore, by counting the average number of maxima in the \textit{un-averaged} cross section in the compound nucleus reaction, or in the electron conductance in chaotic open QD, one obtains the correlation width, also obtainable from Ericson analysis of the
correlation function, Eq. (5). This supplies a powerful double checking of the whole idea of statistical processes in quantum systems, namely being described by a random, S-matrix. \\
The average density of maxima in the case of the external parameter variation, namely $C(\delta X) = \frac{\langle{T^{fl}T^{fl}\rangle}}{[1 + (\frac{\delta X}{X_c})^2]^2}$, is
\begin{equation}
\langle{\rho_X}\rangle = \frac{3}{\pi \sqrt{2} X_c} = \frac{0.68}{X_c}
\end{equation}
The formulae for the average density of maxima can be cast in different forms if the condition for a maximum is changed into a condition on the first and second derivative of the
fluctuating observable. Here one defines the maximum in $T^{fl}(Z)$ as $T^{\prime fl} (Z ) > 0$, and $T^{\prime fl}(Z + \delta Z) < 0$, in the interval $[Z, Z +\delta Z]$, provided $\delta Z$ is
sufficiently small. Expanding in $\delta Z$, the condition for a maximum is then,

\begin{equation}
- T^{\prime\prime fl}(Z) \delta Z > T^{\prime fl}(Z) > 0
\end{equation}

The average density of maxima is then given by integration the joint probability distribution, $P(T^{\prime}, T^{\prime\prime})$, over the random variables,$T^{\prime}, T^{\prime\prime}$, subject to the above constraint.

\begin{equation}
\int_{-\infty}^0 dT^{\prime\prime} \int_0^{-T^{\prime\prime}} dT^{\prime} P(T^{\prime}, T^{\prime\prime}) = - \delta Z \int_{-\infty}^0 dT^{\prime\prime} T^{\prime\prime} P(0, T^{\prime\prime}) \equiv \delta Z \langle{\rho_Z\rangle}
\end{equation}

An appropriate Gaussian distribution function for the the three assumed random variables, $T$, $T^{\prime}$, $T^{\prime \prime}$, which is similar in structure as that of Eq. (), but with a correlation matrix, $C$ given by,

\begin{equation}
 \mathbf{C} = \left(
  \begin{array}{ccc}
  \langle{|T|^2\rangle} & 0 & \langle{T T^{\prime \prime}\rangle}\\
  0 & \langle{|T^{\prime}|^2\rangle} & 0\\
  \langle{T^{\prime \prime} T \rangle}& 0 & \langle{|T^{\prime \prime}|^2\rangle}\\
  \end{array} \right)
\end{equation}

is then constructed. The zero elements in the above correlation matrix are a direct consequence of the correlation function when expanded in powers of $\delta Z$. In fact, the non-zero elements, are obtained from the CF as,

\begin{equation}
\langle{|T^{\prime}|^2\rangle} = - \frac{d^2}{d(\delta Z)^2} C(\delta Z)|_{\delta Z = 0},
\end{equation}
\begin{equation}
\langle{T T^{\prime\prime}\rangle} = \frac{d^2}{d(\delta Z)^2} C(\delta Z)|_{\delta Z = 0},
\end{equation}
and
\begin{equation}
\langle{|T^{\prime \prime}|^2\rangle} = \frac{d^4}{d(\delta Z)^4} C(\delta Z)|_{\delta Z = 0}
\end{equation}.

Since the average density of maxima, Eq. ( ), involves an integral  of only $T^{\prime}$ and $T^{\prime\prime}$, and what at the end we need is $P(0, T^{\prime\prime})$. This distribution is easily obtained from $P(T, T^{\prime}, T^{\prime\prime})$ by integrating over the redundant variable $T$, and setting $T^{\prime} = 0$, getting

\begin{align}
P(0,T^{\prime\prime}) = \frac{1}{2\pi}\frac{1}{\sqrt{\left<[T^{\prime}]^2\right>\left<[T^{\prime\prime}]^2\right>}} \exp{\!\left(-\frac{1}{2}\frac{[T^{\prime\prime}]^2}{\left<[T^{\prime\prime}]^2\right>}\right)}.
\end{align}

One can then obtain the average density of maxima as,

\begin{equation}
\langle{\rho \rangle} = \frac{1}{2 \pi} \sqrt{\frac{\langle{|T^{\prime \prime}|^2\rangle}}{\langle{|T^{\prime}|^2\rangle}}}
\end{equation}

The average density of maxima in the case of energy fluctuations, is then

\begin{equation}
\langle{\rho_{E}\rangle} = \frac{\sqrt{3}}{\pi \gamma}
\end{equation}

and for the variation of the external parameter, where the correlation function is a squared Lorentzian,

\begin{equation}
\langle{ \rho_{X}\rangle} = \frac{3}{\pi \sqrt{2} X_{c}}
\end{equation}

Equations, (38) and (39) coincides with Eqs. (28) and (28), attesting to the equivalence of the two manners of defining a maximum.

The above results are valid for open quantum dots. In the general case of partially open QD. where the tunneling probability (assumed to be equal for all the channels), $\Gamma$, is less than one the formulae for the average densities were recently derived \cite{BHR13}, and they are given by,

\begin{equation}
\langle{\rho_{E}\rangle} = \frac{\sqrt{3}}{\pi \gamma}\sqrt{\frac{9\Gamma^2 - 18\Gamma + 10}{5\Gamma^2 - 10\Gamma + 6}}
\end{equation}

\begin{equation}
\langle{\rho_{X}\rangle} = \frac{\sqrt{3}}{\pi \sqrt{2}X_c}\sqrt{\frac{7\Gamma^2 - 10\Gamma + 6}{2\Gamma^2 - 3\Gamma + 2}}
\end{equation}

The dependence of $\langle{\rho_{E}\rangle}$ and $\langle{\rho_{X}\rangle}$ on $\Gamma$ are exhibited in Figs. (2) and (3).

\begin{figure}
\centering
\includegraphics[width=10cm,clip]{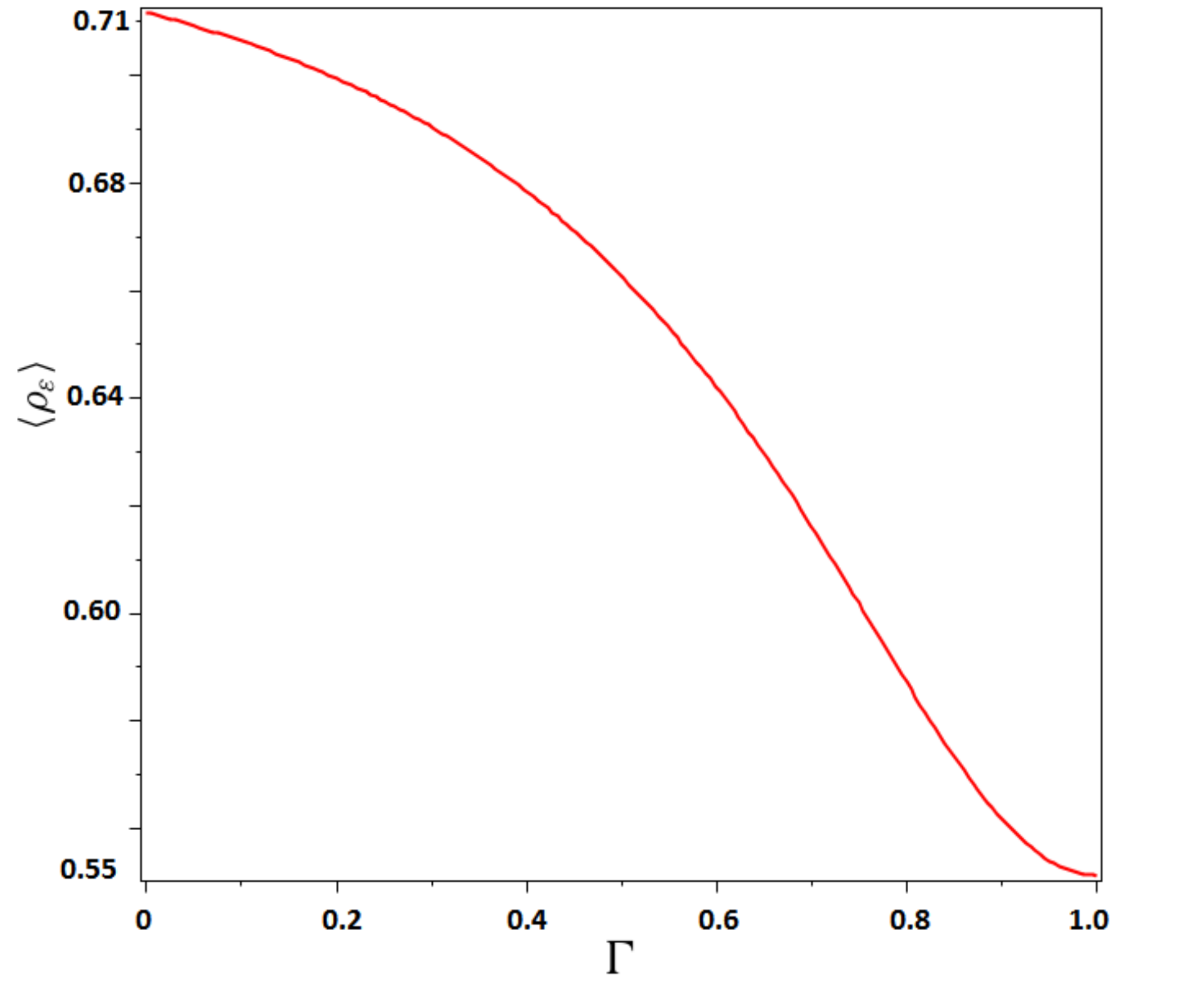}
\caption{The figure shows the average density of maximum dependence on the tunneling rates for parametric variations of energy. Notice the asymptotically decreasing behavior up to the limiting value of 0.55 at $\Gamma$ = 1.}
\label{Fig.2}       
\end{figure}

The above formulae reduce to the former ones in the completely open QD, $\Gamma = 1$. In the opposite limit, $\Gamma = 0$, we get,

\begin{equation}
\langle{\rho_{E}\rangle}  = \frac{\sqrt{5}}{\pi \gamma}
\end{equation}

Surprisingly, for the case of an external parameter variation, we obtain exactly the same result as in the case $\Gamma = 1$,

\begin{equation}
\langle{\rho_{X}\rangle} = \frac{3}{\pi \sqrt{2}X_c}
\end{equation}

Clearly, in this latter case, there is a an extremum (minimum) in the dependence of $\langle{\rho_{X}\rangle}$ on $\Gamma$. This happens for $\Gamma \sim 0.6 $

\begin{figure}
\centering
\includegraphics[width=10cm,clip]{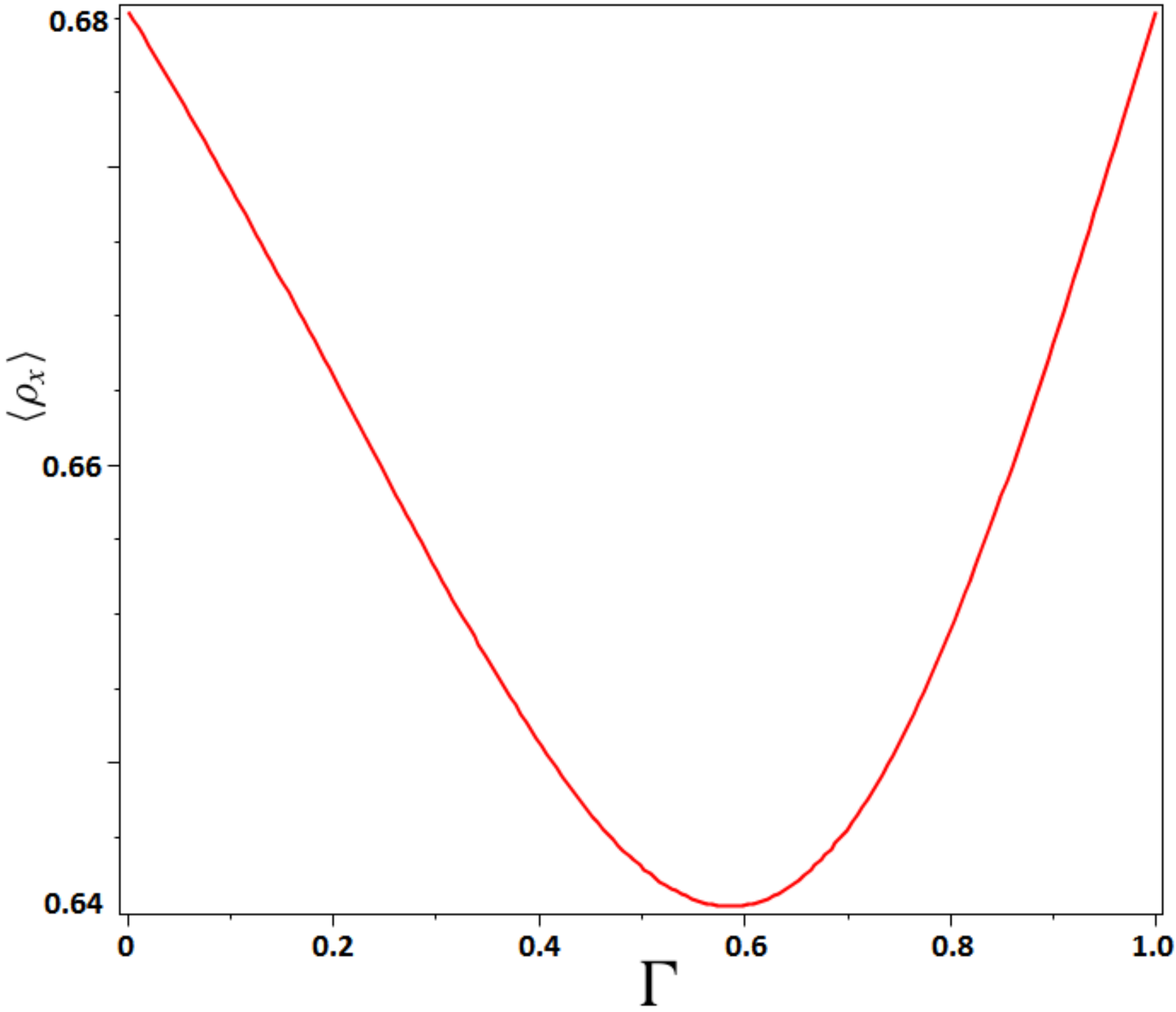}
\caption{The figure shows the average density of maxima dependence on the tunneling rates for parametric variations of a magnetic field. Notice the presence of a minimum and the asymptotically equal values (at $\Gamma$= 1.0, and 0.0).}
\label{Fig.3}       
\end{figure}

\section{Random matrix simulations}

\begin{figure}
\centering
\includegraphics[width=12cm,clip]{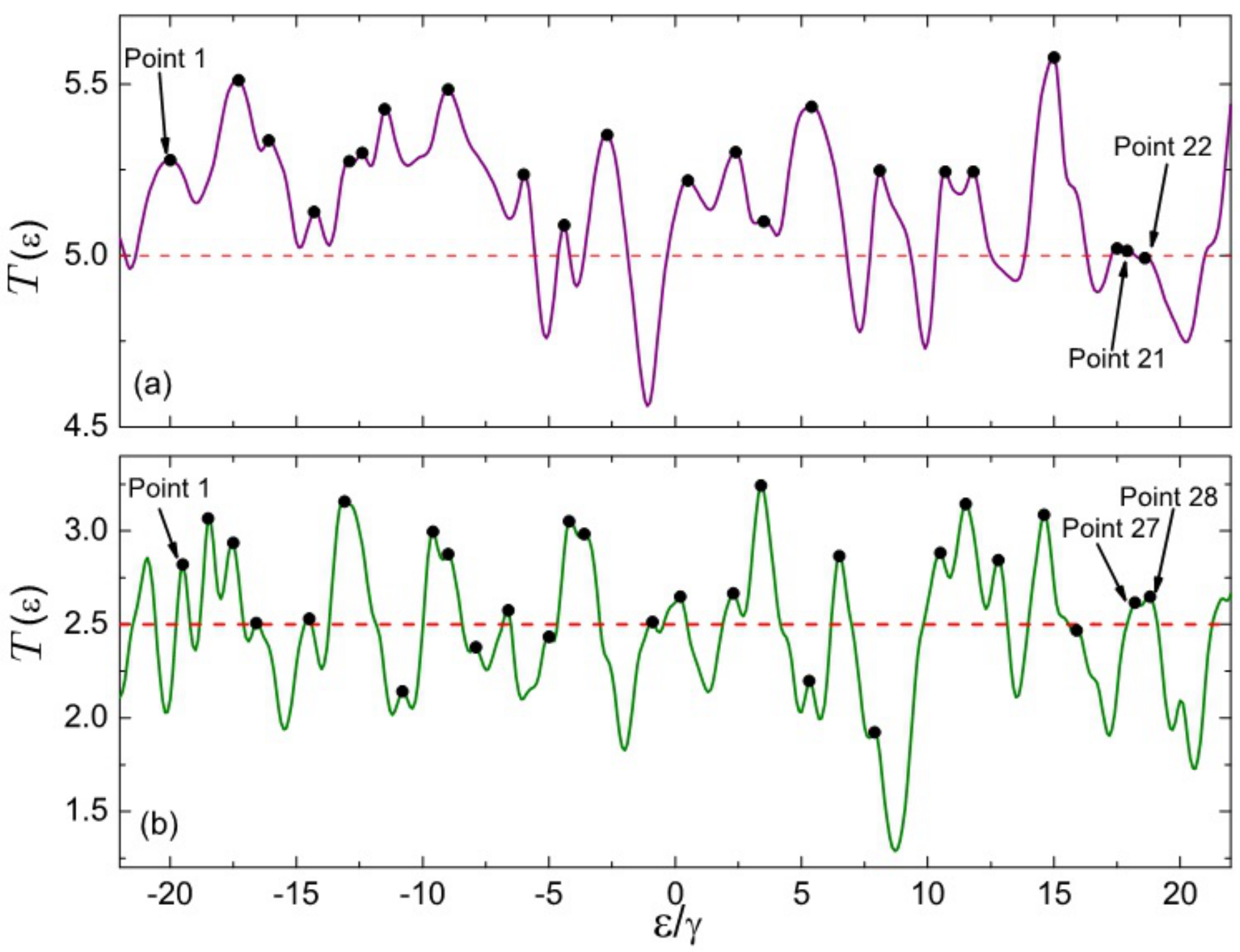}
\caption{Typical dimensionless conductance $T$ as a function of $\varepsilon/\gamma$ for $N=60$ channels coupled to resonances through finite tunneling probabilities. Continuous lines for the numerical results for a single realization of $H$, the dots indicate the maxima of $T$ and the dashed line is the $\varepsilon$-independent conductance average.}
\label{Fig.4}
\end{figure}

\begin{figure}[h!]
	\centering
	        \vskip-0.4cm
		\includegraphics[width= 12cm,clip]{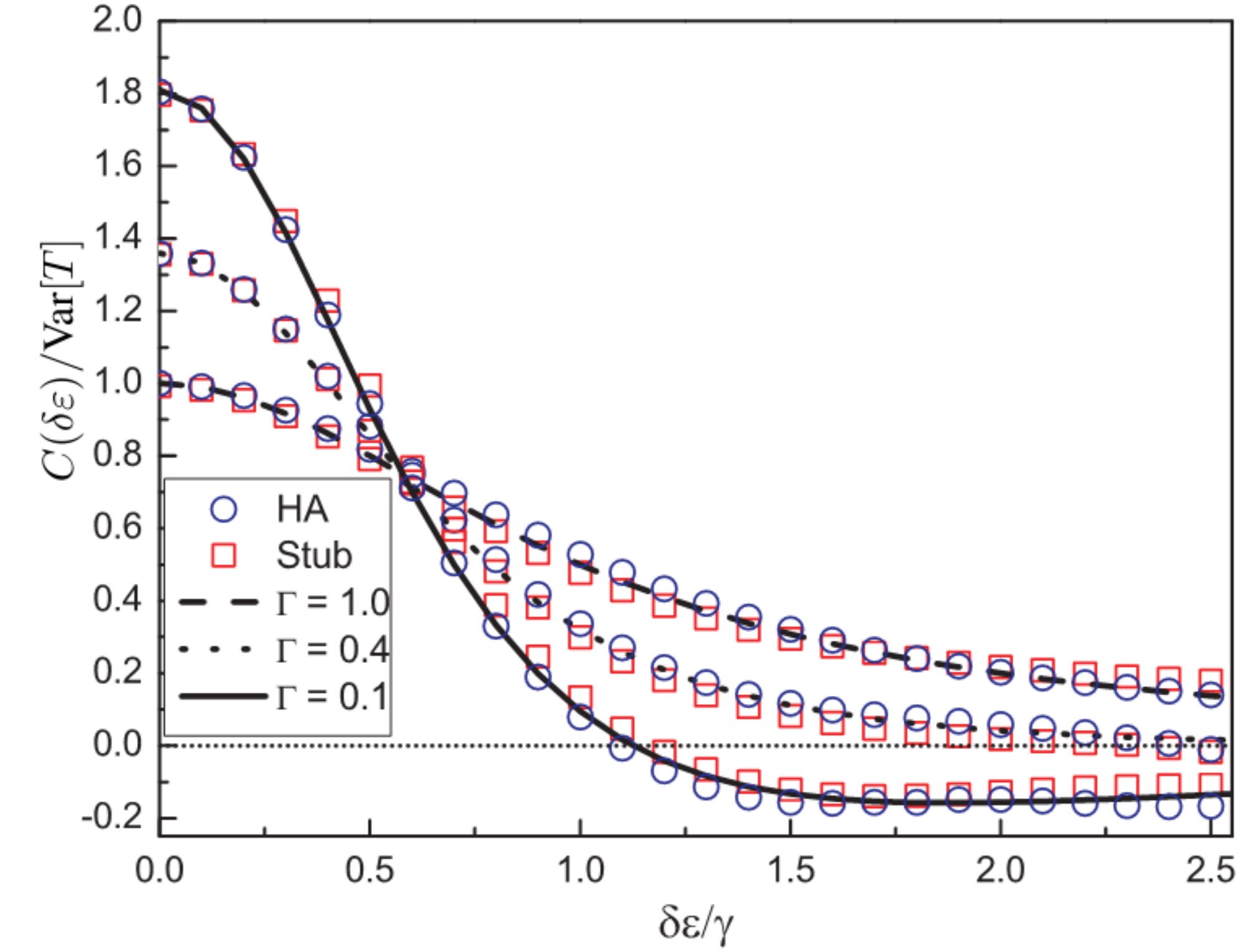}
	\caption{Correlation function (normalized covariance) as a function of parametric variation of energy. Transition from Lorentzian (long dashed line) to a anti-correlation (continuous line) as a function of symmetric tunneling probabilities $\Gamma$. The labels "HA" and "stub" refer to results of simulations performed using the random Hamiltonian model of the S-matrix, Eq. (44), and the stub model of Ref. \cite{brouwer}.}
	\label{Fig.5}
\end{figure}

In this section we present the results of a random matrix simulation based on a random Hamiltonian based S- matrix used extensively by Weidenm\"ulcer and collaborators \cite{weiden1, weiden2, weiden3},

\begin{equation}
S (\varepsilon) = 1 - 2\pi i W^\dagger (\varepsilon - H + i \pi W W^\dagger)^{-1} W,
\end{equation}

where $H$ is a random Hamiltonian matrix of dimension $M \times M$ that describes the resonant states. In the chaotic regime, for which the number of resonances is very large ($M \rightarrow \infty$), we can take $H$ as a member of the Gaussian orthogonal (unitary) ensemble for the symmetric (broken) time-reversal case. The matrix $W$ of dimension $M \times (N_1 + N_2)$ contains the channel-resonance coupling matrix elements. Specifically, for a QD in the chaotic universal regime, the eigenvalues ${\cal W}$ of $WW^{\dagger}$ are connected with the tunneling probability $\Gamma$ through the formula \cite{beenakker97} ${\cal W}=M D (2-\Gamma \pm 2 \sqrt{1-\Gamma})/(\pi^2 \Gamma)$. Since the $H$ matrix is statistically invariant under orthogonal or unitary  transformations, the statistical properties of $S$ depend only on the mean resonance spacing $D$, determined by $H$, and on $W^\dagger W$.

We consider a non-ideal coupling (finite tunneling probabilities), i.e, an arbitrary interaction between the open channels and long-range resonant modes of the QD. We separate the matrix $W$ into blocks corresponding to the respective couplings of QD with the two leads, $W=(W_{1},W_{2})$, where $W_{i}$ is a $M \times N_{i}$ matrix. We disregard the direct processes requiring the orthogonality condition $W_{i}^{\dagger} W_{j}= \omega_{i} M D \delta_{ij}/\pi^2$ with $\omega_{i}=\textrm{diag}(\omega_{i,1},\omega_{i,2},\dots,\omega_{i,N_{i}})$. The tunneling probability (barrier) $\Gamma_{i,c}$ of the channel $c$ and lead $i$ can be written in terms of the diagonal matrix $\omega$ through the relation $\Gamma_{i,c}=\textrm{sech}^{2}(\tau_{i,c}/2)$ with $\tau_{i,c}=-\textrm{ln}(\omega_{i,c})$. We consider equivalent couplings, $\Gamma_{i,c}=\Gamma_{i}$, symmetric contacts, and a large number of resonances in the QD. Without loss of generality, we chose the pure unitary ensemble fo
 r numerical simulation of the previously mentioned Landauer-B\"uttiker conductance CF (identified with the covariance of $T(\varepsilon)=\textrm{Tr}(t^\dagger(\varepsilon)t(\varepsilon))$) using the transmission sub-matrix of Eq.(2 ), which can be written as $t(\varepsilon)=-2 \pi i W_{1}^{\dagger}(\varepsilon - H +i \pi W W^\dagger)^{-1} W_{2}$. The numerical simulation result shown in Fig.[3] shows the conductance obtained through ${\cal N}_r=2000$ realizations of the unitary (complex entries) $H$ matrices with $M=600$ resonances coupled non-ideally with $60$ open channels. The Hamiltonian simulation also exhibits the Weisskopf argument in the ``quasi-closed" limit \cite{weiden4} and nicely confirms an auto-correlation length $\gamma= N\Gamma D/(2 \pi)$. The discrepancies are very small and stay within the statistical precision ${\cal N}_r^{-1/2}$.

We have verified the accuracy of our expressions for the average density of maxima by comparing them with a direct counting of these maxima in Fig.[1]. The results are very good. We have also confirmed the accuracy of the expressions for the correlation functions $C(\varepsilon)$ and $C(X)$ using the results of the random matrix simulations, Fig.[\ref{Fig.2}]. Again, the results are very good, and confirm the correctness or our analytical results.

\section{Discussion and conclusions}

In this contribution we have argued that the statistical behavior of open Quantum Dots, resembles that of the compound nucleus. The average density of maxima in the fluctuating part of the electron conductance is directly related to correlation length which characterizes the correlation function. Simple formulae are obtained for this density in the general case of a partially open QD, with tunneling probabilities less than unity. We have compared our analytical results with random matrix simulations carried out for the $S$-matrix. Our findings can be useful both in compound nuclear reactions and in QD research. Further application of our results to the chaotic behavior of electron conductance in graphene flakes \cite{flakes1, flakes2, flakes3} is underway. The results obtained here for the case of partially open chaotic QD, can be potentially useful in compound nucleus reaction research. The limit $\Gamma = 1$ refers to overlapping resonances (strong absorption), while $\Gamma \sim 0$ corresponds to isolated resonances (weak absorption). In fact \cite{Alhassid} presented the data for the reaction p + $^{35}$Cl $\rightarrow$ $\alpha$ + $^{32}$S, which shows the anti-correlation effect. We are currently pursuing this line of research.\\

\bigskip

\end{document}